\documentclass[psfig,epsfig,12pt]{article}
\usepackage{graphicx}
\begin{document}
\baselineskip 0.2in
\begin{center}
\begin {title}
\large {\bf Punctuated Equilibrium and Power Law in Economic Dynamics}
\end{title}
\end{center}
\smallskip
\begin{center}
Abhijit Kar Gupta
\end{center}

\begin{center}
{\it Department of Physics, Panskura Banamali College, Panskura, East Midnapore, Pin Code: 721152, West Bengal, India}\\
{\em e-mail:}~{kg.abhi@gmail.com}
\end{center}

\bigskip
\noindent{\bf Abstract:}

\smallskip
An interesting toy model has recently been proposed on Schumpeterian economic dynamics by Thurner {\it et al.} \cite{thurner} following the idea of economist 
Joseph Schumpeter \cite{schump}. Punctuated equilibrium dynamics is shown to emerge from this model and some detail analyses of the time series indicate SOC kind of behaviours. 
The focus in the present work is to toss the idea whether the dynamics can really be like a self organized critical (SOC) type. This study indicates that it is necessary to 
iccorporate the concepts of 'fitness' and 'selection' in such a model in the line of the biological evolutionary model by Bak and Sneppen \cite{bs} in order to obtain power law and thus 
SOC behaviour.

\bigskip
\noindent{\bf Introduction:}

\smallskip

The time series from non-equilibrium dynamics where random and rapid (or chaotic) events are 
followed by a relatively longer period of inaction or
'stasis' (eventless plateau) can be termed as 'punctuated equilibrium' as proposed by Gould and Eldrege \cite{gould} through a 
series of works (papers and books) in connection with Darwinian evolution. 
In the famous Bak-Sneppen model (BS) \cite{bs} and in subsequent other such toy models, the idea of punctuated equilibrium had been explored 
and also power laws were shown to appear in different meaningful outcomes. The occurrence of punctuation in the kind of evolutionary time series 
may be understood in terms of
the fact that a system gets a periodic or probabilistic kick or stirring (randomly or selectively) from time to time to come out of a 'plateau'. However, to 
obtain a power law, one perhaps needs to connect this disturbance (stirring) to the dynamical state of the system at any point of time. In other words, there has 
to be a feedback in the system. In absence of a feedback or control on the dynamics, the outcome may not necessarily be a power law. The distributions of some relevant quantities may turn into a Poissonian or a kind of exponential one.
An exponential distribution implies that the consecutive events would follow each other at a more or less regular time 
intervals where the very long waiting times are not likely to appear.  
Power law distributions imply slow decay with fat tails and that 
can be ascribed to the occurrence of rare events. The rare events often happen with a burst, followed by the long waits. So the underlying 
process may be related to some kind of delay or decision making or waiting. In biological terms, there are selection and extinction 
followed by new species via mutation. 

So, our idea is that the presence of a kind of feedback or control on the dynamics drive the outcome towards a power law. If we examine the BS model of 
evolution, there is the concept of selection and then extinction of the 'species' (a site) with lowest fitness which is 
replaced by another with new fitness value. There is also the concept of mutation implemented through the upgrading of the fitness value of the
nearby species (or sites). (The BS model was on a one dimensional lattice though. So the concept of mutation was easily implemented there.) 
The design of algorithm makes it possible to obtain power laws in the distribution of avalanche sizes, in waiting time distributions
or in the growth of a particular 'species' (site).  

With the above view, it would be interesting to look at
the dynamics of social interactions or evolution of market (or products or technologies) which can essentially be captured in terms
of competition, selection, extinction etc. The present toy model as proposed recently by Thurner {\it et al.} \cite{thurner} (We refer this 
as 'Thurner model' henceforth for convenience.) conceptualizes
the essential ideas of Schumpeterian economic dynamics (See an illuminating brief 
discussion by Mark Buchanan \cite{buchanan}.). Importantly, 
punctuated equilibrium results in here along with some interesting features. 
Our main interest is to look at the dynamics and to understand how or whether a SOC type criticality results in
with such dynamical rules or with what kind of dynamical rules.

\bigskip
\noindent{\bf The Model:}

\smallskip

As conceived by Schumpeter, the economic activity 
can be treated as the appearance and disappearance of goods and services that are linked to technological innovations.   
New goods appear in the market and displace the old ones (This is 'creative destruction' as coined by 
Schumpeter.). In reality, this keeps on happening when a new technology (or a new concept) is invented with the help of two (or more) existing technologies (or concepts). 
A new product appears in the market which in turn produce a cascading of goods or products and that may in 
effect influence a series of products to be abolished from the market. 
In the algorithm of Thurner model \cite{thurner} on Schumpeterian dynamics, the existing goods or products (or old technologies) combine to create a 
new one (new technology). 

In more precise terms, the model has a provision for $N$ products, each of these can either exist or does not exist 
at a given time. The state of each product at a certain time changes via creation or annihilation. So a state vector $\sigma_i(t)$ is 
associated with a product $i$ which has two components: 1 (exist) or 0 (does not exist).
A product 'does not exist' means either it is eliminated from the market or it is not yet created. Next it is considered that 
a certain product gets created by the 
combination of two existing products chosen at random. 
Thus at a certain time, a number of products exist and one can keep track of the total count over a suitably defined time 
scale. This total count keeps changing with time as it is obvious from the scenario of random creation and destruction. The average 
value of this total count, ${1\over N}\Sigma\sigma_i(t)$ (like the magnetization in spin system) is termed as 'product diversity'. It is
now interesting to look at the time series of product diversity.

The product diversity varies randomly over a period of time around some equilibrium value (dynamic equilibrium) and then eventually halts to be stuck in
a plateau (as demonstrated in \cite{thurner}) as if in a 'stasis'!. 
A kind of stirring in the system is required from time to time to kick the system out of a plateau. An 'innovation' or an appearance of a radically new product or a new kind of 
technology can trigger that. 
Considering that, the system starts varying randomly all over again in a cluster 
before reaching to another plateau. So there are intermittent plateaus or 'stasis' in the so called time series. Therefore, a kind of 
punctuated equilibrium emerges from the dynamics. 

In this model, a probabilistic innovation brings new products (or 
technologies) in the market forcing a old one to fade from time to time. In algorithmic terms, this is implemented by flipping the state of a product: some products are born and some other get abolished from the market.
However, the abolition of a product seems to have no connection to the state of 
the dynamics in the Thurner model. In our view, a feedback at this stage is needed. It will be interesting to see how a feedback mechanism does 
influence the outcome of the dynamics.

Now, borrowed the idea from Bak-Sneppen (BS) model, we like to check how a critical punctuated 
equilibrium may appear in this Schumpeterian dynamical model with the control on the dynamics through fitness. In BS model, the extinction of a species is done 
with that of lowest fitness (The fitness is competitive.). In a sense, the concept of fitness acts as a marker on the system which 
can be thought to control the dynamics in some way. If the extinction is done arbitrarily, the power law would not probably appear (this can easily be checked numerically
in BS model). 
Therefore, in support of this view, it has been thought to implement some feedback through some marker in the so called 
Thurner model. This is in a way, selection and extinction which certainly have an impact over the dynamics for the next 
time step (can be thought of as a feedback). In fact in reality, this happens. Products or technologies with 
lowest efficiency (or fitness) eventually get abolished from the market. 
So in the above algorithm in Thurner model, the rule $\# 2$ is required to be modified to see a real 
SOC type behaviour while retaining rule $\# 1$.
We thus incorporate a new step towards this in the well stated algorithm in \cite{thurner}.

The dynamics in \cite{thurner} is implemented in the following way. 
The creation of a product $k$ at time $t$ may or may not be possible from two products $i$ and $j$. 
This is determined by predesigning a creation tensor $\alpha_{ijk}^{+}$, the action of this is in general the following:.  
$\sigma_k(t+1)=\alpha_{ijk}^{+}\sigma_i(t)\sigma_j(t)$
The matrix elements are considered to be either 0 or 1. When each of the elements on the right is 1, the left side 
equals 1 which means a creation of a product. 

As the new product may in turn drive another existing product out of the market, there will be an
annihilation tensor whose action is given by the following:
$\sigma_k(t+1)=1 - \alpha_{ijk}^{-}\sigma_i(t)\sigma_j(t)$
Here the annihilation is possible when the element of annihilation tensor, $\sigma_{ijk}^{-}=1$ and also individual ingredients,
$\sigma_i(t)=\sigma_j(t)=1$.
Whether a product will exist or not, will depend upon the net effect of creation and destruction by all the possible
pairs of products in the system. In this way, when all the products are updated (sequentially or randomly), we call 
that to be one time step. 
Next, the flipping of the state of a randomly chosen product was done in the original work with some probability $p$.

\bigskip
\noindent{\bf The Algorithm:}
\smallskip

$\# {\bf 1.}$ $\Delta$ = $\sum_{ij}$($\alpha_{ijk}^{+}$$-$$\alpha_{ijk}^{-}$)$\sigma_i(t)\sigma_j(t)$, 

\indent for $\Delta > 0$, $\sigma_k(t+1)=1$; $\Delta < 0$, $\sigma_k(t+1)=0$ and for $\Delta=0$, no change.

\smallskip

$\# {\bf 2.}$ $\sigma_l(t+1)=1-\sigma_l(t)$ with some probability $p$. Flipping of the state

\bigskip
\noindent To incorporate {\em feedback}, we introduce the following {\bf new rule $\# 2$:}
A fitness parameter $f_i(t)$ is assigned to each product. In the new rule $\# 2$: the product with the lowest fitness value at time $t$ is abolished {\it i.e.}, we set $\sigma_k(t+1)=0$ for that product for the next time 
step. The creation of a radically new product is done randomly as before, {\it i.e.}, $\sigma_j(t+1)=1$ with probaility $p$ for some $j$.   
In the numerical simulation, the fitness value is assigned randomly (random number between 0 and 1) to each 
product to start with and whenever a product is created by rule $\# 1$.
    
\bigskip

\bigskip
\noindent{\bf The Simulation Results:}
\smallskip

The dynamics emerges with interesting punctuated equilibrium pattern (as originally obtained in Thurner model) where there are fluctuations and intermittent plateaus in the 
time series of product diversity ($\sum\sigma_i(t)$). To examine the character of this dynamics (whether it is of SOC as in \cite{bs}), we decided to check some distributions.
For this we specifically looked at the time-plateau (or waiting time) size distribution.
For the dynamics to be critical, this distribution should follow a power law as the emergence of power 
law can be associated with a critical or self organized dynamics. Some simulation results are reported here. 

\begin{figure}[h]
\begin{center}
\includegraphics[height=3.7in, width=4.5in]{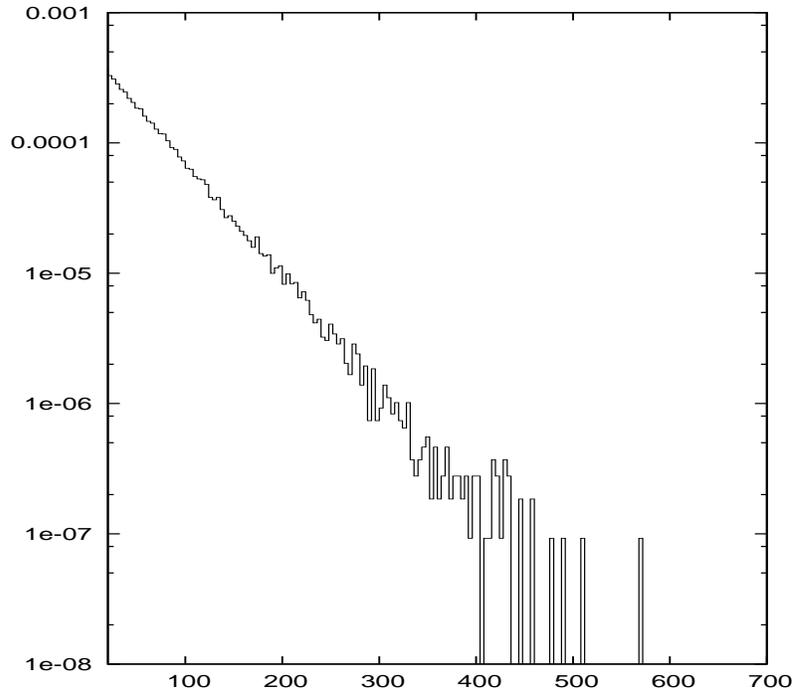}
\caption{Waiting time (plateau size) distribution ($P(\tau)$ vs. $\tau$) from the model in \cite{thurner}. The exponential law is
evident from the semi-log plot. This distribution is made from $3\times 10^6$ time steps. In this case 
simulation results are obtained for innovation probability $p=0.0002$ and the maximum number of products $N = 100$.}
\end{center}
\end{figure}

The entries in the matrices for the creation and destruction operators are 0 and 1. In the numerical simulations, the outcome could be altered by designing the 
creation and destruction matrices and choosing the value of the probability $p$. However, this does not affect the final results and thus the main concluson (the 
distribution and all that).

\begin{figure}[h]
\begin{center}
\includegraphics[height=3.7in, width=4.5in]{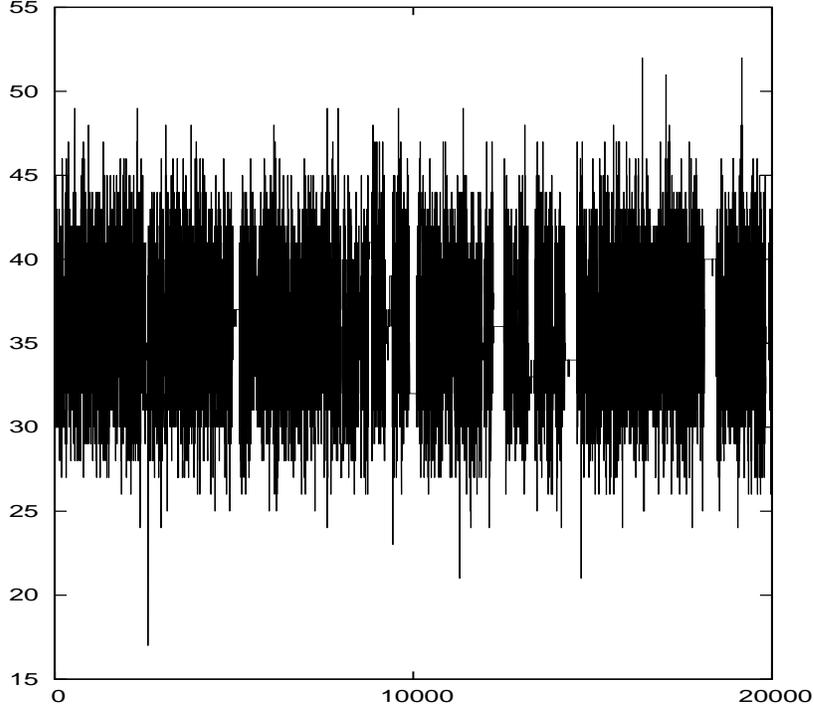}
\caption{A typical punctuated equilibrium appearing in the time series of total product count with the new rule in place. 
Data is obtained from the simulation for innovation probability $p=0.0002$ and the maximum number of products $N = 100$.}
\end{center}
\end{figure}

We analyse the time series (for a certain value of $p$) for product diversity, ${1\over N}\Sigma\sigma_i(t)$ and looked at the time plateau (waiting time) distribution in that.
For a comparison, we first examine the time series from Thurner model (with rules $\#1$ and $\#2$ as stated above). It can be seen from {\bf fig.1} that the distribution is clearly 
an exponential one. Therefore, it appears that the  dynamics may not correspond to a critical or SOC behaviour. To check further, we obtained the cumulative returns for a 
particular product over time. It appears like a devil's stair case type plot (not shown here). The
distribution of plateau size from this also shows an exponential character (results not shown here). So
it can be said that this present Schumpeterian dynamics is perhaps a non critical punctuated equilibrium. In this context, it may be said that
the punctuated equilibrium with non critical dynamics occurs in nature as reported in \cite{non-critical} for snow avalanches, The snow avalanches arrive in clusters and bursts 
with rapid changes over short intervals and interrupted with periods of non action (stasis) but the system is claimed to be non critical.
Therefore, it is in general important to understand the dynamics producing punctuated equilibrium with or without criticality. 

\smallskip
\begin{figure}[h]
\begin{center}
\includegraphics[height=3.7in, width=4.5in]{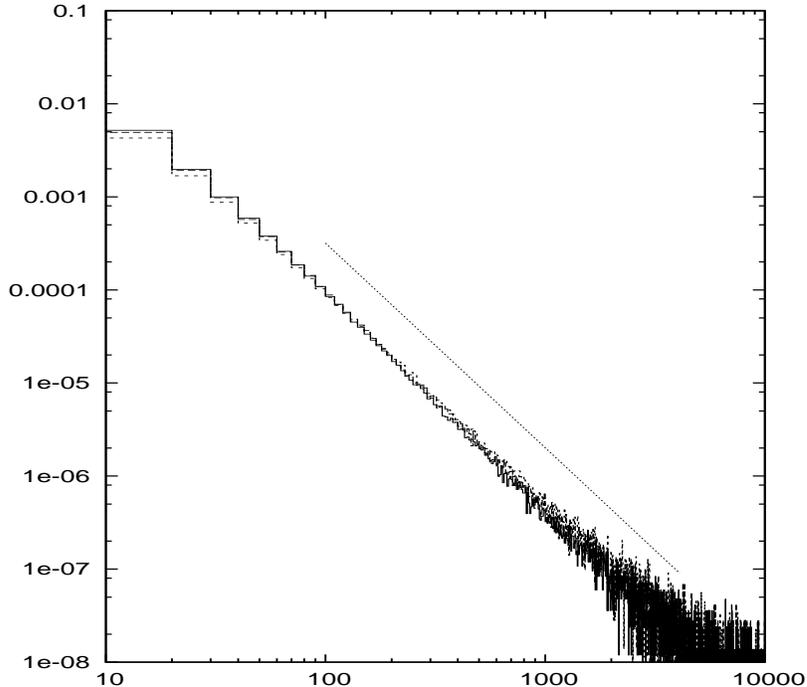}
\caption{Waiting time (time-plateau size) distributions with the new rule in place for three different values of 
$p$ = 0.0002, 0.0003 and 0.0005. Here, for quick results we considered $N=50$. The power law is
evident from the log-log plot and the straight line shows the slope of -2.2. Each of the distributions is made
from around $7\times 10^6$ time steps.
}
\end{center}
\end{figure}


The time series for total count is plotted in {\bf fig.2} to show the kind of punctuated equilibrium.
The waiting time distribution obtained from this time series is clearly a power law as is evident from {\bf fig.3}.
We also examined the evolution of a single product. In the case of SOC, the successive returns (cumulative value) can be 
seen to be a devil's staircase. We have examined that and this appears to be so in our simulation. Here also, the 
waiting time ($P(\tau)$ vs. $\tau$) distribution for the evolution of a single product emerges as a power
law (The result is not shown here.)

In conclusion, we may say that this is only a preliminary observation related to the model by Thurner {\it etal.} on economic dynamics. 
Further work is in progress. It may be mentioned that in recent past, a number of simple toy models for evolution (biological and other) have been 
proposed in the literature to understand some interesting features like the punctuated equilibrium and self organization pertaining 
to complex systems. While trying to interprete the outcomes of the resultsi in the present toy model, it has been 
observed that the criticality and power law is possibly achieved if there is a selection or decision or 
delay that is linked to the dynamics of the system itself. In the Bak-Sneppen model, the concept of feedback was 
implemented through the selection, extinction and followed by mutation. In other words, it may be said whenever 
there is a kind of feedback is in place to control the dynamics, power law or self organization may be achieved. 
Thinking on that line, a similar idea of feedback (or selection and extinction) has been implemented through a 
'fitness' parameter in the model of economic evolution. The idea of mutation could not however be implemented as the model is not on a 
lattice and thus the interactions between the products are
arbitrary. Interestingly, some years ago 
Barabasi \cite{barabasi} came up with a very simple model to establish an idea where it was shown that 
decision based queuing process plays a crucial role in bursty nature and power law in human dynamics.
Therefore, in the light of all those, it has been an attempt to add a necessary step in the present Thurner model 
for which SOC type critical behaviour clearly emerges in addition to interesting punctuated equilibrium. 
More detailed numerical works and mathematical analysis are needed to confirm the present view and to 
understand the dynamics in a substantial way.

\bigskip
\noindent {\bf Acknowledgments:}
\smallskip

The author is thankful to Abdus Salam ICTP, Trieste for the kind hospitality during the visit of one month where the work was carried out. 
This work is partly supported by the UGC minor grant No. PSW-86/06-07. The author likes to thank {\it D. Stauffer}, {\it M. Marsili} and {\it M. Chatterjee} for their opinions
and suggestions regarding this work.

\end{document}